\documentclass[a4paper]{article}
\usepackage{amssymb}
\newtheorem{thm}{Theorem}
\newtheorem{clm}{Claim}
\newcommand{\lozzy}{\lozenge_{l^\mu}^l}

\title{A constructive proof presenting languages in $\Sigma^P_2$ 
that cannot be decided by circuit families of size $n^k$.}
\author{Sunny Daniels \\ e-mail: sdaniels@lycos.com}
\begin{document}
\maketitle
\begin{section}{The long and complicated history of this result}
I proved this result in 1998 when I was a doctoral student at Rutgers
University in New Jersey; the lecturer in charge of the Complexity
Theory course, Professor Eric Allender, presented a non-constructive
proof of this result, but said that, at the time, he did not know of
any constructive proof of the result.  He promised an ``instant A'' in
the course (198:538 Complexity Theory) to any student who could
constructively prove this result.

I devised this constructive proof of the result at the time.  As far
as I know, I was the only student in the class to succeed in
constructively proving this result.  I wrote up a draft of the result,
nearly in form suitable for publication, printed it out and presented
it to Allender: It was eight pages long, including the cover sheet.
There were a few minor mistakes, which I corrected by hand on the
printed-out result, and an incomplete reference to some chapters
by Allender, Louis and Regan in the {\em CRC Handbook of Algorithms
and Theory of Computation}.  However, because of various personal
circumstances at the time, I did not have the opportunity to attempt
to publish it.

I left the USA and moved to the UK in 1999, and, again, because of
personal circumstances (I went through a period of financial
hardship), did not have the opportunity to attempt to publish it.  In
2001 I was fully occupied with an IT job in the UK unrelated to
Complexity Theory, and hence did not have the time to attempt to
publish the result.  At that time, I incorrectly assumed that I did
not have a copy of the result with me; I had left behind several
suitcases of personal possessions with friends in New Jersey, and I
wrongly assumed that I had left all copies of the result in New
Jersey.

However, I managed to obtain a Council flat in the UK in November 2002
(I was again unable to find a job, and wanted the financial security
of living in social housing), and so had more financial stability,
but, by this time, I had almost completely forgotten about the result
that I proved in 1998 in New Jersey; I was a bit occupied with looking
for work, and carrying out some unpaid research in areas unrelated to
Complexity Theory.

Then, in January 2004, I became curious about the fate of this result;
I e-mailed Allender, and was told that the same result had been proved
by Cai and Watanabe and presented at COCOON
2003\protect\cite{cocoon03}; apparently a copy of my result had been
faxed to the COCOON 2003 conference and shown to Cai and Watanabe, who
agreed to acknowledge it in a subsequent publication.  It appears that
Cai and Watanabe's COCOON 2003 paper was also published as technical
report C-161 at Tokyo Institute of Technology\protect\cite{c161}.  I
do not remember getting any response from Cai and Watanabe for several
years after this; in 2008, however, in response to a bit of pressure
to rewrite my CV, I e-mailed Allender about this again.  It appears
that Cai and Watanabe responded by withdrawing technical report C-161
and replacing it with report C-256\protect\cite{c256}.  Technical
report C-256 acknowledges my result.

However, Allender claimed to have only ever had a hard-copy of the
result (he e-mailed me a copy of the faxed image in 2004, if I
remember rightly), and so the only way that I could have published it
properly at the time would have been to re-typeset the whole result,
which seemed to be a bit tedious; the mediocre resolution of the faxed
image only compounded the problem.

I ultimately moved back to New Zealand (I originally graduated from
Massey University in 1993 and then the University of Auckland in 1998)
in November 2008, but was still unaware of the fact that I had the
\LaTeX\  source of the result until a few weeks ago (in August 2014).

I have five Iomega Zip cartridges of data from my time as a student at
Rutgers, but, since I ceased to have easy access to an Iomega Zip
drive when I left Rutgers in 1999, and I had never owned a Zip drive
until a few weeks ago, I was unaware of the fact that one of the Zip
cartridges contained the \LaTeX \ source for my 1998 result.  The fate
of the \LaTeX\  source of my 1998 result was still unclear to me
until a few weeks ago.

I received an Iomega Zip drive (for free) from a programmer, Ed
Church, here in Auckland, a few weeks ago.  I installed it in one of
the old Linux computers that I use, and looked at the contents of the
five cartridges, expecting to find some interesting old files, but
certainly not expecting to find my 1998 result!  I was amazed to find
the complete \LaTeX \ source of the old 1998 result on one of the Zip
cartridges.  (Note added later: Since finding the original 1998 result
on a Zip cartridge, I found a mostly-complete attempt to re-typeset
the result, dated 21 July 2005, in a collection of files copied from
the hard drive of the computer that I had in the UK at the time.  I
now vaguely recall corresponding with Allender about attempting to
publish the result at the time, but I don't clearly remember why I did
not complete this attempt to re-typeset the result at the time; I
remember getting involved in an unpaid IT project for a local business
at about that time).

My result from 1998 (the version from the {\em original} \LaTeX \
source) is presented here, with the minor mistakes corrected, and the
references fixed.
\end{section}
\begin{section}{Acknowledgements}
Thanks to Professor Eric Allender for originally making me aware of
this, at the time, unsolved problem back in 1998.  Thanks for his
assistance with my apparent previous attempt to publish it in 2005.

Thanks to Professors Jin-Yi Cai, at the University of Wisconsin at
Madison, and Osamu Watnabe at Tokyo Institute of Technology for
acknowledging this result in 2008.  Thanks also to Cai and Watanabe
for giving this result a name (in the bibloigraphy of
\protect\cite{c256}), which I have used as inspiration for the title
of this article: my original unpublished manuscript in 1998 had no
title.  (I feel that my title more accurately reflects the exact form
in which I state my result in section \ref{main_result}).  And thanks
finally to Ed Church, a programmer here in Auckland, for giving me the
Iomega Zip drive which enabled me to re-discover and retrieve the
original \LaTeX \ source of the result.
\end{section}
\begin{section}{The Result to be Proved}
\begin{thm}
\label{main_result}
Take any $k \in \mathbb{N}$. 
Then there is a language $L_k \in \Sigma^P_2$ (with
$L_k \in \{0,1\}^*$) that is not decidable by any
circuit family of size $n^k$.
\end{thm}
{\em Proof:}
We will define $L_k$ in terms of two other languages
 $\Gamma$ and $\Lambda_k$.
We will define the language $\Gamma$ in section \ref{gamma_def}.
We will then define the language $\Lambda_k$ in section
\ref{lambda_def}, after we have made some necessary preliminary
definitions.
\end{section}
\begin{section}{The first $q$ strings of length $n$:}
For any $n \in \mathbb{N}$, it is obviously possible to list
all of the strings over $\{0,1\}$ in increasing lexicographic order.
For example, if $n = 3$, the following is a list of all of the strings
over $\{0,1\}$ of length $n$ in increasing lexicographic order:
\[
  000, 001, 010, 011, 100, 101, 110, 111
\]
Now, for any $q \in \{ 0, 1, \ldots, 2^n \}$, we can form
the set of the first $q$ strings from this list of all of the strings
in $\{0,1\}^n$.  Let $\lozenge_q^n$ denote the set formed in this way.
For example:
\[
  \lozenge_5^3 = \{000, 001, 010, 011, 100 \}
\] 
\end{section}
\begin{section}{The language $\Gamma$:}
\label{gamma_def}
Let $\Gamma$ be the language defined by:
\[
  \Gamma = \left\{ 0x  
	\left| \begin{array}{ll}
		x \in \{ 0, 1 \}^* \\
       		\textrm{$x$ represents a satisfiable Boolean expression} 
	\end{array} \right.
  \right\}
\]
Here, we assume the existence of some intuitively-simple 
method of representing Boolean expressions by bit strings.
In section \ref{bool_bin}, we will give a particular example of
such a mapping.  We will then assume that the mapping used in
the definition of $\Gamma$ is the mapping given in section
\ref{bool_bin}.
Now, it is a well-known fact that the satisfiability problem is
in NP, so there is a non-deterministic polynomial-time Turing Machine
$M_\alpha$ to solve it.  It is obviously possible to modify the
machine $M_\alpha$ to discard the initial zero before starting the 
task of solving the satisfiability problem.
Hence, there is a non-deterministic polynomial-time (and hence
certainly $\Sigma^P_2$) Turing machine $M_\beta$ that accepts
$\Gamma$.
\end{section}
\begin{section}{A problem that can be solved with a 
non-deterministic Turing Machine and a satisfiability circuit}
\label{reduc_to_SAT}
Let $B^`$ be any language over the alphabet $\{0,1\}$
in which all of the strings are of the same length
 $n \in \mathbb{N}$.  Then define the language ${K}$ as follows:
\[
  K = \left\{
    ( 1^n, 1^m, B^` ) \left|
      \begin{array}{l}
	n, m \in \mathbb{N} \\
	n \leq m \\
	B^` \subseteq \lozenge_m^n \\
	\textrm{$\exists$ $n$-input circuit $C^`$ of size $m$, 
		$L(C^`) \cap \lozenge_m^n = B^`$}
      \end{array}
    \right.
  \right\}
\]
Here, $B^`$ is represented in the input string as a comma-delimited list
of all of the strings in $B^`$.  We claim that:
\begin{clm}
\label{K_KRT_SAT}
The language $K$, as defined above, is Karp-reducible to the
satisfiability problem.  (A definition of Karp-reducibility is
given by, for example, Allender et al\protect\cite{chap28}).
\end{clm}
Now, it should be fairly obvious that the following algorithm
is in NP and accepts $K$, and hence $K \in \mathrm{NP}$:
\begin{enumerate}
\item
Check that $n \leq m$.  If we find that $n \not\leq m$,
we reject the input.
\item
Run through all of the strings in $B^`$, checking
that each one is in $\lozenge_m^n$ (i.e. checking
that each one is of length $n$ and lexicographically precedes
the binary representation of $m$).
If we find some string $B^`$ to not be in $\lozenge_m^n$,
we reject the input.
\item
Existentially guess a circuit $C^`$ of size $m$.
\item
Run through all of the strings $x \in \lozenge_m^n$,
checking that, for each such string $x$, $x \in B^`$ 
if and only if $C^`$ accepts $x$.
If we find some string $x \in \lozenge_m^n$ that
does not satisfy this condition, then we reject the
input.  Otherwise, we accept the input.
\end{enumerate}
Now, there is a well-known result called {\em Cook's Theorem} that
states that every language in NP is Karp-reducible to the
satisfiability problem.  For a proof of Cook's Theorem, see, for
example, Allender et al\protect\cite{chap28}.  The proof of Cook's
Theorem given by Allender, et. al\protect\cite{chap28} is
constructive; it explicitly gives a procedure for constructing the
``transformation function'' $\tau$ referred to in the definition of
Karp-reducibility.\footnote{I called the transformation function
$\tau$ in my original 1998 version of this result: I presume that this
was the name for the transformation function used in lectures.
However, it appears that Allender et al's article\protect\cite{chap28}
uses the name $f$ for the transformation function.  I prefer to stick
with my original name, $\tau$, for the transformation function.} 
Part of the definition of Karp-reducibility
implies that some polynomial function $p(|x|)$ is an upper bound
on the length of $\tau(x)$, for any string 
 $x \in \{0,1,{\bf (},{\bf )},{\bf ,}\}^*$ of
length $n$.  But, since we assume that $n \leq m$, the lengths
of the strings $1^n$, $1^m$ and $B^`$\footnote{i.e. the
comma-delimited list of all elements in $B^`$} are all polynomial
in $m$.  Hence, the length of $x$ is also polynomial in $m$.
Since the composition of two polynomial functions is also a polynomial
function, this implies that there is a polynomial function $q$ for
which $q(m)$ is an upper bound on the length of $\tau((1^n,1^m,B^`))$,
for any $(1^n,1^m,B^`) \in K$.  Since $q$ is a polynomial function, there is
some $\delta \in \mathbb{N}$ for which $q(m) \leq m^\delta$ for
all $m \geq 2$.
\end{section}
\begin{section}{Recursive Definition of the Satisfiability Problem}
\label{recur_sat}
In this section, we will discuss a property of the satisfiability
problem that will later be important to our proof.  In order to do
this, we will start by giving a formal definition of the
satisfiability language SAT.  This language is defined to be the
language over 
$\Sigma_B = \{0,1,{\bf (},{\bf )},{\bf \vee},{\bf \wedge},{\bf v}\}$ for which,
for any $x \in \Sigma_B^*$:
\begin{enumerate}
\item
Boolean expressions are constructed from the symbols in $\Sigma_B$.
 ${\bf \vee}$, ${\bf \wedge}$, ${\bf (}$ and ${\bf )}$ denote
disjunction, conjunction and precedence in the usual way.  Variables
are denoted by strings of $0$s and $1$s following the symbol ${\bf v}$
(e.g. ``${\bf v}101$'', ``${\bf v}11001$'', ``${\bf v}0011$'').
Literals are denoted by $1$ and $0$, for ``true'' and ``false'',
respectively.
For any $x \in \Sigma_B^*$ that is {\em not} a well-formed Boolean
expression, $x \not\in \textrm{SAT}$.
\item
For any $x \in \Sigma_B^*$ that represents a well-formed Boolean
expression, $x$ is defined to be in SAT if and only if this Boolean
expression is satisfiable.
\end{enumerate}
Now, given this definition, and the intuitively-obvious properties of
the satisfiability problem, it should be clear that SAT is the unique
language $Q$ over $\Sigma_B$ that satisfies the following three
properties\footnote{This is actually a special case of a property of
languages known as {\em self-reducibility}.  For a definition of
self-reducibility, see, for example, Ko\protect\cite{selfred} 
(Note added in 2014: I don't clearly remember exactly which
defintion of self-reducibility I was using in 1998: it appears that
I incorrectly assumed that it was defined in Allender et 
al\protect\cite{chap28} at the time.  Presumably I was using a
definition given in lectures.  However, Ko\protect\cite{selfred}
gives several different precise definitions of self-reducibility,
including {$d$-self-reducibility}, which he claims SAT to satisfy.
It appears that $d$ in this context stands for ``disjunctive'').}
\begin{enumerate}
\item
For any $x \in \Sigma_B^*$ that {\em does not} represent a well-formed
Boolean expression, $x \not\in Q$.
\item
For any $x \in \Sigma_B^*$ that represents a well-formed Boolean
expression with no variables, $x \in Q$ if and only if $x$ evaluates
to true.
\item
For any $x \in \Sigma_B^*$ that represents a well-formed
Boolean expression with variables, let $v_1$ be the variable in $x$ that has the
lexicographically-first name.  Let $x_0$ be the expression resulting
from substituting the literal ``$0$'' for every occurrence of $v_1$ in
 $x$.  Similarly, let $x_1$ be the expression resulting from
substituting the literal ``$1$'' for every occurrence of $v_1$ in $x$.
Then $x \in Q$ if and only if either $x_0 \in Q$ or $x_1 \in Q$.
\end{enumerate}
\end{section}
\begin{section}{Binary representation of Boolean expressions:}
\label{bool_bin}
In section \ref{recur_sat}, we represented Boolean expressions as
strings over the alphabet:
\[
  \Sigma_B = \{ 0, 1, (, ), \vee, \wedge, {\bf v} \}
\]
Clearly, then, Boolean expressions can be represented as bit strings
just by replacing each of the symbols in $\Sigma_B$ by a different
three-bit string.  For example, we could use the following
mapping: 
\begin{center}
\begin{tabular}{cc}
  Symbol in $\Sigma_B$ & Replacement Bit String \\
  \hline
  $0$ & $000$ \\
  $1$ & $001$ \\
  $($ & $010$ \\
  $)$ & $011$ \\
  $\vee$ & $100$ \\
  $\wedge$ & $101$ \\
  ${\bf v}$ & $110$ \\
\end{tabular} 
\end{center}  
Note that the three-bit sequence $111$ does not represent any symbol
of $\Sigma_B$.  This allows us to use a string of $1$s to pad out the 
binary representation of any given string over $\Sigma_B$ to any
desired length (whether or not the desired length is a multiple of three).

We will assume, from here on, that this method of representing Boolean
expressions is used in the definition of $\Gamma$ in section
\ref{gamma_def}.
\end{section}
\begin{section}{Upper bound on number of circuits of given polynomial
size}
\label{ub_poly_size}
We will refer to the number of gates in a given circuit as the
{\em size} of that circuit.  Now, fix any natural number $\eta$.
Now, take any natural number $n$, and consider the task of
constructing an arbitrary $n$-input circuit of size at most $n^\eta$.
There are clearly at most $(n^\eta)^2 = n^{2\eta}$ different
(source,destination) pairs for wires in the circuit.
Hence, a circuit can be described by:
\begin{enumerate}
\item
Choosing one of the $2^{n^{2\eta}}$ different subsets of the
(source,destination) pairs for the wires in the circuit.
\item
Choosing, for each of the $\leq n^\eta$ non-input gates,
one of a finite number $c$ of types ({\em and}, {\em or},
{\em not}, etc) for the gate.  Clearly there are at most
$c^{n^\eta}$ different ways in which this can be done.
\end{enumerate}
Therefore, the total number of ways in which the circuit
can be constructed is at most:
\begin{eqnarray*}
  c^{n^\eta}2^{n^{2\eta}} 
	& \leq & c^{n^{2\eta}}2^{n^{2\eta}} \\
	& =    & (2c)^{n^{2\eta}} 
\end{eqnarray*}
Hence, there is clearly some natural number $\mu$
for which, for all $n \geq 2$, there strictly fewer than $2^{n^\mu}$
ways of constructing a circuit of size $n^\eta$.
\end{section}
\begin{section}{The language $\Lambda_k$:}
\label{lambda_def}
Let $k$ be any given natural number.
Clearly there is some natural number $\eta$ for which
 $n^\eta \geq (n + 1)^k$ for every natural number $n \geq 2$.
Let $\mu$ be the number $\mu$ derived from $\eta$ in the way
described in section \ref{ub_poly_size}.
Now, let $\Lambda_k$ be the language accepted by a
 $\Sigma^P_2$ Turing machine $M_{\Lambda_k}$.
The machine $M_{\Lambda_k}$ is defined to operate as follows:
\begin{enumerate}
\item
\label{step_1}
Check that the input string is of the form $1x$,
where $x$ is a string of binary digits.
If the input string is not of this form, then halt and reject.
Clearly this step can be performed in time linear (and hence
polynomial) in the input length.  Let $l$ denote the length $|x|$
of $x$.
\item
\label{step_2}
Existentially guess a circuit $G$ of size $(l^{\delta \mu} + 1)^k$ with
 $l^{\delta \mu}$ inputs.  Here, $\delta$ is as defined in 
section \ref{reduc_to_SAT}, under the assumption that $m$ is 
always equal to $l^\mu$.  Since $\delta$, $\mu$ and $k$ are all
independent of $l$, and the input length to $M_{\Lambda_k}$ is
 $l + 1$, then the amount of time taken to existentially guess $G$
is polynomial in the length of the input to $M_{\Lambda_k}$.
\item
\label{step_3}
Existentially guess a subset $B_l$ of $\lozzy$.
Since $\lozzy$ contains $l^\mu$ strings, each of length $l$,
the amount of time required for this existential guessing step
is proportional to $ll^\mu$, which is polynomial in $l$.
\item
\label{step_4}
Perform a four-outcome universal choice operation.  Depending upon the
outcome of the universal choice, branch ahead to either step
\ref{outcome_1}, step \ref{outcome_2}, step \ref{outcome_3} or step
\ref{outcome_4}.
\item
\label{step_5}
\label{outcome_1}
Universally guess a circuit $C$ of size\footnote{We
regard the {\em size} of a circuit as the number of gates that it
contains.} $(l + 1)^k$.
This universal guessing step will clearly require time at most
proportional to $(l + 1)^k$ to guess the types (and, or, not, etc) 
of the gates, plus an additional time at most proportional to 
 $(l + 1)^{2k}$ to determine the pairs $A$, $B_l$ of gates for which
to run a wire from $A$ to $B_l$.
Hence, the time required for this universal guessing step is at most
polynomial in the size of the input to $M_{\Lambda_k}$.  
\item
\label{step_6}
Check that, for the circuit $C$, 
 $L(C) \cap \lozzy$ is not equal to $B_l$.
To do this, all we have to do is run through all of the strings in
 $\lozzy$, and check that one of them is
either in $B_l$ but not in $L(C)$, or in $L(C)$ but not in $B_l$.
Now, $\lozzy$ contains $l^\mu$ strings,
each of which is of length $l$.  So computing all of the strings
in $\lozzy$ requires time at most proportional to $ll^\mu$.
But, for each string $y \in \lozzy$, we perform two checks:
\begin{enumerate}
\item
Check whether $y \in B_l$.  If we represent $B_l$ as the list of all of
its elements, then we can use sequential search to check whether or
not $y \in B_l$.  Since $B_l$ is a subset of $\lozzy$, then the list
representation of $B_l$ will be of length at most $ll^\mu$.
Hence, the time required for the search will be at most polynomial
in $l$.
\item
Check whether $y \in L(C)$.  This just requires the circuit $C$
to be simulated for the input $y$.  Clearly this can be done
in time at most proportional to the number of wires in $C$,
i.e. at most proportional to the square of the size of $C$.
Since the size of $C$ is at most proportional to $(l + 1)^k$,
the square of the size of $C$ is at most proportional to 
 $(l + 1)^{2k}$, which is polynomial in $l$.  
\end{enumerate}
So, this step involves a polynomial number of checks,
each of which takes polynomial time.  So the total time required
for this step is polynomial in the size of the input to $M_\Lambda$.
If it turns out that 
 $L(C) \cap \lozzy$
is not equal to $B_l$, then we halt and accept.  If, on the other hand,
 $L(C) \cap \lozzy = B_l$, then we halt and reject.
\item
\label{step_7}
\label{outcome_2}
Universally guess a subset $B^`$ of $\lozzy$ that
lexicographically precedes $B_l$.  Since $\lozzy$
contains $l^\mu$ strings, each of which is of length $l$, the
amount of time required to perform this universal guessing operation
is proportional to $ll^\mu$ (clearly one pass through $B_l$ and
 $B^`$ is sufficient to determine that $B^`$ lexicographically
precedes $B_l$).  
\item
\label{step_8}
Take the deterministic polynomial-time Turing Machine for computing the
function $\tau$ mentioned in section \ref{reduc_to_SAT}.  Simulate the
execution of this Turing Machine on the input $(1^l,1^{l^\mu},B^`)$.
(We established in section \ref{reduc_to_SAT} that the function
$\tau$ can be computed by a deterministic Turing machine in time
polynomial in {\em its} input length.  Since $B^` \subseteq \lozzy$,
it is clearly also true that the size of the argument to $\tau$ is
polynomial in the length of the input to $M_{\Lambda_k}$.
Hence, this simulation can be performed in time polynomial in
$M_{\Lambda_k}$'s input length.
Since $l \leq l^\mu$ and $B^` \subseteq \lozzy$,
 $\tau((1^l,1^{l^\mu},B^`))$ will be satisfiable if and only if there is
some $n$-input circuit $C^`$ of size $m$ for which 
 $L(C^`) \cap \lozzy = B^`$.
Now, since the second argument to $\tau$ is $1^{l^\mu}$,
the properties of $\tau$ shown in section \ref{reduc_to_SAT}
show that $l^{\delta \mu}$ is an upper bound on the length
of $\tau((1^l,1^{l^\mu},B^`))$.  Hence, it is possible to 
evaluate the circuit $G$ on the input $\tau((1^l,1^{1^\mu},B^`))$
(if $\tau((1^l,1^{1^\mu},B^`)$ is a string of less than 
 $l^{\delta \mu}$ bits, then we pad it out to $l^{\delta \mu}$ bits
by adding 1s to the end of it.) 
If $G$ accepts this input, then we halt and {\em
reject}.  If $G$ rejects this input, then we halt and accept.
(Clearly we can evaluate the circuit $G$ on this input in polynomial time).
\item
\label{step_9}
\label{outcome_3}
Check that $G$ computes the language SAT for its given input length.
By the result in section \ref{recur_sat}, all we need to do in order
to do this is to:
\begin{enumerate}
\item
Universally guess a bit string $y$ whose size equals the number of inputs
of $G$.
\item
If $y$ does not represent a well-formed Boolean expression, see
 whether $G$ rejects $y$. (If $G$ rejects $y$, halt and accept.
 Otherwise halt and reject.)
\item
If $y$ represents a well-formed Boolean expression with no variables,
then halt and accept if $y$ evaluates to true, halt and reject if $y$
evaluates to false.
\item
If $y$ represents a well-formed Boolean expression with variables,
then try substituting $0$ for the lexicographically-first variable
in $y$ and evaluating $G$ on the resulting expression.  
Also try substituting $1$ for the lexicographically-first variable
in $y$ and evaluating $G$ on the resulting expression.
Check that $G$ accepts $y$ if and only if $G$ accepts at least
one of the expressions resulting from these substitutions.
(If so, halt and accept.  If not, halt and reject).
\end{enumerate}
Clearly, the amount of time required to perform this check is at
most polynomial in $G$'s input length, which is at most polynomial in
$M_{\Lambda_k}$'s input length.  
\item
\label{step_10}
\label{outcome_4}
Check whether $x \in B_l$.  If so, halt and accept.  Otherwise halt and
reject.  (Obviously, from the construction of $B_l$, this check can be
performed in time polynomial in $M_{\Lambda^k}$'s input length.)
\end{enumerate}
\end{section}
\begin{section}{Definition of $L_k$ itself}
Now, we define (for each $k \in \mathbb{N}$) $L_k$ to be the union of
 $\Gamma$ and $\Lambda_k$.  It is now trivial to construct a
 $\Sigma^P_2$ machine to accept $L_k$: this machine just looks at the
first bit of the input string, and runs the machine for $\Gamma$
or $\Lambda_k$ accordingly.
\end{section}
\begin{section}{Conclusion of Proof}
Now, assume for the sake of a contradiction that Theorem
\ref{main_result} is false.  Hence, for some $k \in \mathbb{N}$, $L_k$
is decidable by a circuit family of size $n^k$.
\begin{subsection}{Deriving a Satisfiability Circuit}
\label{sat_exists}
Then, we can obviously hardwire the first input of each circuit in the
family to $0$, and thereby obtain a circuit family of size $(n+1)^k$
that decides whether a given input bit string represents a satisfiable
Boolean expression.

Now, as in section \ref{lambda_def}, let $\eta$ be a natural number
for which $n^\eta \geq (n + 1)^k$ for every natural number $n \geq 2$.
Let $\mu$ be the number $\mu$ derived from $\eta$ in the way described
in section \ref{ub_poly_size}.
Then, for every natural number $l \geq 2$, we can set 
 $n = l^{\delta \mu}$ and thereby conclude that 
there is a circuit of size $(l^{\delta \mu} + 1)^k$
with $l^{\delta \mu}$ inputs that decides if the input bit string
represents a satisfiable Boolean expression. 
\end{subsection}
\begin{subsection}{Deriving the $\mho_k$ circuit}
Also, by taking each circuit in the circuit family for $L_k$, and
hardwiring its first input to $1$, we can obtain a circuit family of
size $(n+1)^k$ that decides the language $\mho_k$, defined by:
\[
  \mho_k = \{ x \in \{0,1\}^* : 1x \in \Lambda_k \}
\]
Now, consider the result of running the machine $M_{\Lambda_k}$
on the input $1x$, for an arbitrary bit string $x \in \{0,1\}^*$.
Obviously step \ref{step_1} will not reject the input string $1x$,
so execution will pass through to step \ref{step_2}.

Now, look at steps \ref{step_2}, \ref{step_3} and \ref{step_4}.
From section \ref{sat_exists} above, we know that there {\em is} a
circuit of size $(l^{\delta \mu} + 1)^k$ that solves the
satisfiability problem for inputs of size $l^{\delta \mu}$.
Furthermore, one of the places that step \ref{step_4} universally branches ahead
to is step \ref{step_9}.  But step \ref{step_9} halts and rejects
unless the circuit $G$ solves the satisfiability problem.
Hence, we can assume that the circuit $G$ existentially guessed in
step \ref{step_2} solves the satisfiability problem for all inputs of
size $l^{\delta \mu}$.

Now, consider the subset $B_l$ of $\lozzy$ existentially guessed in step
\ref{step_3}.  We have already established that $l^\eta > (l + 1)^k$
for every $l \geq 2$.  Then, by the result in section
\ref{ub_poly_size}, there are strictly fewer than $2^{l^\mu}$
ways of constructing a $l$-input circuit of size $l^\eta$, and hence certainly
fewer than $2^{n^\mu}$ ways of constructing an $l$-input circuit $E$ of size 
 $(l + 1)^k$.  Hence, there are strictly fewer than $2^{l^\mu}$
different sets $L(E) \cap \lozenge^l_{l^\mu}$ that can be constructed
from such $l$-input circuits $E$.  Hence (since 
 $2^{|\lozenge^l_{l^\mu}|} = 2^{l^\mu}$), there must be at least one subset
of $\lozenge^l_{l^\mu}$ that does not equal 
 $L(E) \cap \lozenge^l_{l^\mu}$ for any $l$-input circuit $E$ of size
 $(l + 1)^k$; let's call every such subset of $\lozzy$ a {\em
non circuit-constructible} subset.  
Now, clearly the check in steps \ref{step_5} and
\ref{step_6} ensures that the machine $M_{\Lambda_k}$ will halt and
reject unless the set $B_l$ guessed in step \ref{step_3} is
non circuit-constructible.  Now, our assumption that the circuit
$G$ solves the satisfiability problem clearly implies that steps 
\ref{step_7} and \ref{step_8} have the effect of
checking that $B_l$ is the lexicographically-first
non-circuit-constructible subset of $\lozzy$; step \ref{step_8} halts
and rejects unless $B_l$ is the lexicographically-first such set.
Hence, we can assume that the set $B_l$ existentially guessed in step
$\ref{step_3}$ is the lexicographically-first non
circuit-constructible subset of $\lozzy$.  But then clearly
step \ref{step_10} has the effect of making $M_{\Lambda_k}$ halt and
reject if and only if $x$ is in the lexicographically-first non
circuit-constructible subset of $\lozzy$.  Hence, the language
$\mho_k$ defined at the beginning of this section is
the set of all strings $x \in \{0,1\}^*$ for which $x$ is in the
lexicographically-first non-circuit-constructible subset of $\lozzy$
(where $l = |x|$, but $\mu$ is the same for all $x$).
This, together with our earlier definition of
non-circuit-constructibility,
implies that $\mho_k$ is not decided by any circuit family of size
$(n+1)^k$.  This is a contradiction, since we established earlier in
this section that $\mho_k$ {\em does} have a circuit family of size
$(n+1)^k$.  

Therefore, we conclude that theorem \ref{main_result} is true.
\end{subsection}
\end{section}
\bibliographystyle{alpha}
\bibliography{my_theorem}
\end{document}